\documentclass[a4paper]{jpconf}
\usepackage{graphicx,bm}
\newcommand{\BB}{$\bm B$ }
\newcommand{\HH}{$\bm H$ }
\newcommand{\MM}{$\bm M$ }

\begin{document}

	\title{Thermodynamics of magnetized dense neutron-rich matter}
	
	\author{J P W Diener$^{1,2}$}
	
	\address{$^1$Department of Physics and Astronomy, Private Bag 16, Botswana International University of Science and Technology, Palapye, Botswana}
	\address{$^{2}$Centre for Higher and Adult Education, Faculty of Education, Stellenbosch University, Matieland, South Africa}
	
	\ead{dienerj@biust.ac.bw}

\begin{abstract}
A neutron star is one of the possible end states of a massive star. It is compressed by gravity and stabilized by the nuclear degeneracy pressure. Despite its name, the composition of these objects is not exactly known. However, from the inferred densities, neutrons will most likely compose a significant fraction of the star’s interior. While all neutron stars are expected to have a magnetic field, some neutron stars (``magnetars'') are much more highly magnetized than others with inferred magnetar surface magnetic field is between $10^{14}$ to $10^{15}$ gauss.
While neutron stars are macroscopic objects, due to the extreme value of the stars’ energy, pressure, and magnetic field the thermodynamics on the microscopic scale can be imprinted on the star’s large scale behaviour. 
This contribution focusses on describing the thermodynamics of magnetized dense neutron matter, its equation of state and to explore conditions of a possible ferromagnetic state, contributions from the magnetized vacuum, as well as possible observational implications.
\end{abstract}

\section{Introduction}
Neutron stars are a generic term for compact stellar objects with masses up to a couple of times that of the Sun, but with radii of the order of ten kilometres, corresponding to densities similar to nuclear matter and beyond \cite{annMR}.  Central to our understanding of neutron stars is the equation of state (EoS) of nuclear matter \cite{annnuc}. Neutron stars are observed as a variety of astrophysical objects \cite{nsobs,magobs} and the category of interest here is the class of neutron stars known as magnetars.  Magnetars are highly magnetized neutron stars with inferred surface magnetic fields of between $10^{14}$ to $10^{15}$ gauss (and even greater in the magnetar interior)\cite{magnetars}.  At these magnetic field strengths the magnetic interaction could be strong enough to influence nuclear interactions and therefore EoS of the interior. As with other neutron stars, the magnetar EoS is expected to, at least in part, consist of nuclear matter.  Unfortunately there are currently no direct probes of the neutron star/magnetar EoS, thus EoS properties and characteristics are inferred from neutron star observations as well as nuclear theory, experimental results and predictions \cite{annnuc}. \\
\\
Magnetars are observed to exhibit a variety of interesting behaviour including sudden spin ups or spin downs of the rotation frequency (called ``glitches'' and ``anti-glitches'') as well as X- or $\gamma$-ray bursts \cite{magnetars}.  Furthermore, recently there have been indications that magnetars have been associated with the enigmatic and yet unexplained powerful fast radio bursts (FRBs) \cite{FRBmag}.  FRBs are highly energetic, seemingly once-off, bursts of radio waves of extragalactic origin that supersede all other previous detections of radio bursts in terms of energy (see \cite{FRBmag} and references there-in).\\
\\
For the purposes of this contribution I will assume the magnetar interior to consists of magnetized charge-neutral and beta-equilibrated nuclear matter.  This system has been studied previously (see \cite{DienerSpin} and references there-in) and the contribution here expands by including an additional coupling constant parameter set and discussion of observational consequences.
To motivate these predictions, a short overview of the details of the model and its implications will be given. 
%
%
%
\section{Formalism}\label{sec:form}
A generalized magnetized gas of non-interacting fermions is described by the following Lagrangian density:
\begin{eqnarray}
	{\cal L}&=&
	\bar{\psi}{ (x)}
	\left[
	i\gamma^{\mu}\partial_{\mu}-q_f\gamma^\mu A_\mu
	-\frac{g_f}{2}F^{\mu\nu}\sigma_{\mu\nu}-m
	\right]\psi{ (x)}
	-\frac{1}{4}F^{\mu\nu}F_{\mu\nu}	\label{calL}
\end{eqnarray}
is considered, 
where 
$\psi$ is the fermion field operator, $A^\mu$ the gauge field, $F_{\mu\nu} = \partial_\mu A_\nu- \partial_\nu A_\mu$ is the electromagnetic field tensor, and $\sigma^{\mu\nu} = \frac{i}{2}\left[\gamma^\mu,\gamma^\nu\right]$ are the generators of the Lorentz group  \cite{DienerSpin}.\\
\\
This description can easily be extended for interacting nuclear systems within the context of relativistic mean-field models, where mesons mediate the nuclear interactions.  Parameter sets NL3 \cite{NL3},  FSU \cite{FSU}, IU-FSU \cite{IUFSU}, and FSU2 \cite{FSU2} denoting different optimized values of the coupling strengths were used.  The IU-FSU and FSU2 are of particular interest as these parameter sets were optimized for dense matter and neutrons applications \cite{IUFSU,FSU2}.\\
\\
In Eq.\ (\ref{calL}) two couplings to the magnetic field are included, one for the charge, $q_f$, of the fermion and another for the coupling of the fermion's magnetic dipole moment, $g_f$.  For a fundamental particle, such as an electron, the $g_f$ coupling will make the anomalous contributions to the electron dipole moment, but for composite fermions (baryons) like protons and neutrons, $g_f$ take into account the higher-order contributions to the baryon dipole moments due to their internal structure. Despite being a misnomer \cite{DienerPhD}, the $F^{\mu\nu}\sigma_{\mu\nu}$-coupling is referred to as the ``anomalous magnetic moment'' or ``AMM''-coupling \cite{Brod00}.\\
\\
%
%
%
%
%
Using the Heaviside-Lorentz unit convention (natural units with $\epsilon_0 = 1$) the magnetic field will be described in terms of the relationship
\begin{eqnarray}\label{HBM}
{\bm H} = {\bm B} - {\bm M}\mbox{ or }{\bm B} = {\bm H }+ {\bm M},
\end{eqnarray}
where \MM is the magnetization of the Fermi gas in response to the magnetic field, \HH is the magnetic field that is externally applied to the system, and \BB is the magnetic field experienced by the fermions through the couplings with $A^\mu$ and  $F_{\mu\nu} $ in Eq.\ (\ref{calL}) \cite{DienerSpin}. \\
\\
Assuming a form for $A_\mu$ so that ${\bm B}=B\hat{z}$, the spectrum of the neutral fermions and charged fermions are 
\begin{eqnarray}
\omega(|{\bm k}|=k,\lambda)
&=& \sqrt{\left(\sqrt{k_{\bot}^2+{m}^2}+\lambda g_f B\right)^2+k_{z}^2} \label{singlepatN},
\end{eqnarray}
and 
\begin{eqnarray}
\omega(k_z,\lambda,n)=\sqrt{k_z^2+\left(\sqrt{{m}^2+2\,\alpha\,q B n}+\lambda  g_f B\right)^2}\label{singlepatP},
\end{eqnarray}
where 
	$\lambda = \pm 1$ distinguishes the different orientations of the fermion dipole moment, 
	$k_{\bot}^2 = k_{x}^2+k_{y}^2$ the momenta perpendicular to $\bm B$,
	$k_z$ is the momentum in the $z$-direction,
	$\alpha =\mbox{sgn}(q_fB)$, and
	$n$ labels the Landau levels \cite{DienerPhD,Brod00}.
The thermodynamic quantities of the system described by Eq.\ (\ref{calL}) for 
temperature $T$, volume $V$, and chemical potential $\mu$ are related through 
\begin{eqnarray}
	\Omega(B,\mu, T) = -\frac{\mbox{ln}\, {\cal Z}}{V \beta},
\end{eqnarray}
where $\beta =(k_B T)^{-1}$ the inverse temperature times Boltzmann's constant, and $ \cal Z$ the grand canonical partition function of the system.  For $T=0$,  
	\begin{eqnarray}\label{Ofulla}
	\Omega(B,\mu) &=& \Omega_f(B,\mu)+\Omega_{EM}(B)\\
	&=& \sum_\lambda \int\frac{d^3 k}{(2\pi)^3}
	\big\{
	-\omega+
	\left(
	\omega-\mu
	\right)
	\,\Theta\!
	\left[
	\,\omega-\mu
	\right]
	\big\}
	+\frac{B^2}{2}.\label{Ofullb}
	\end{eqnarray}
Thus $\Omega(B,\mu) $ has contributions from the magnetized fermions, $\Omega_f(B,\mu)$, and the free magnetic field, $\Omega_{EM}(B)$. $\Omega_f(B,\mu)$ also contains a contribution from the vacuum, the first term in the integrand in (\ref{Ofullb}).  However, the magnetized vacuum contribution will be ignored as it has been found to make a minimal contribution \cite{DienerVac} and we will proceed with the no-sea approximation. \\
\\
The magnetization, $|{\bm M}|=M$, in Eq.\ (\ref{HBM}) is related to $\Omega(B,\mu) $ through
 \begin{equation}\label{M}
 	M=-\left(\frac{\partial\Omega_f}{\partial B}\right)_{\mu,T}.
 \end{equation}
 Furthermore, the (thermodynamic) pressure in the system is given by $P=-\Omega$.  Due to the breaking of the spherical symmetry of the Fermi surface, as evident from Eqs (\ref{singlepatN}) and (\ref{singlepatP}), the pressure also becomes directional with $\bm B$, with the longitudinal $P_{\parallel}$ and transverse $P_{\perp}$ pressures are given by \cite{DienerSpin,Bland82,Ferrer2010,Strick} 
 \begin{eqnarray}
 P_{\parallel} &=& -\Omega = -\Omega_f-\frac{1}{2}{B}^2\label{paraP}\\
 P_{\perp} &=& -\Omega_f-{B M}+\frac{1}{2}{B}^2 = P_{\parallel}+B(B-M)\label{perpP}
 \end{eqnarray}
which induces an anisotropic in the  pressure of
 \begin{eqnarray}\label{aniso}
 P_{\perp}-P_{\parallel}  =  B(B-M)=BH.
 \end{eqnarray}
 The configuration of $P_{\perp}$ and $P_{\parallel}$ in terms of $B$ or $H$ is illustrated in figure \ref{fig:anisob}.
 %
 %
 %
 \section{Magnetic field configurations}\label{sec:mag}
 \subsection{$H\neq 0$}
  From Eq.\ (\ref{HBM})  if $H\neq0$, then the measure of anisotropy in $P$ depends on 1) the magnitude of $H$, which will have to be significant, as well as 2) the size (as well as sign) of the magnetic response of the system/$M$.  Hence the anisotropy in the pressure of the magnetized gas depends on the external applied magnetic field $H$ as well as the characteristic of the system's magnetic response.  Furthermore $H\neq0$ is not the lowest energy/equilibrated state of the system, since minimizing $H$ minimizes the Gibbs Free Energy of the system \cite{DienerSpin}.  Therefore, when applying the Lagrangian in (\ref{calL}) to describe a physical system, careful consideration needs to be given to the physical parameters and energy considerations imposed on the system.  
 \subsection{$H=0$}
\begin{figure}[h]
	\begin{minipage}{16pc}
		\centering
		\includegraphics[width=10pc]{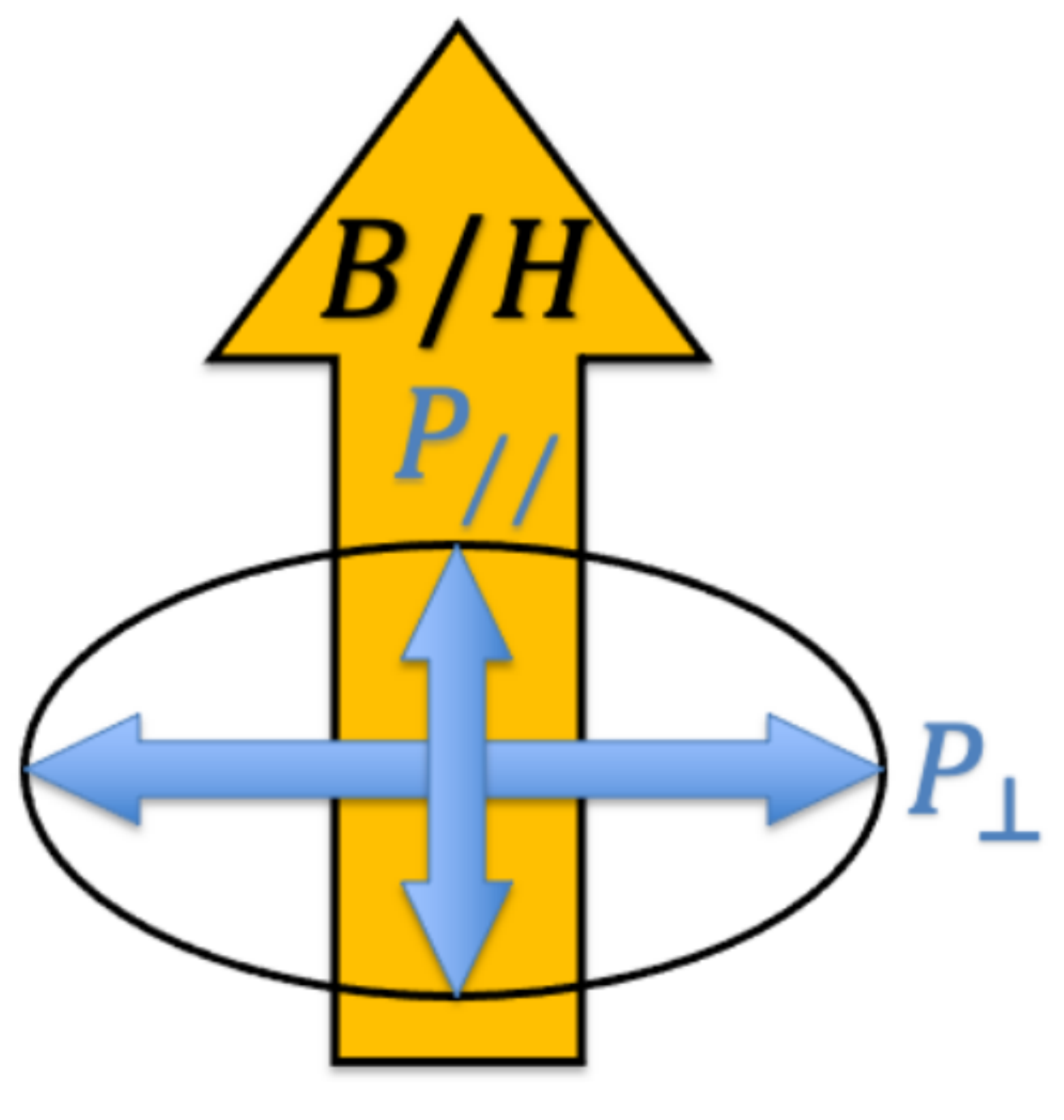}
		\caption{\label{fig:anisob}Illustration of the relative orientation of $P_{\perp}$ and $P_{\parallel}$ with regards to the magnetic field.  It is worthwhile to remember that for most observed neutron stars the rotational axis and the magnetic field axis is not aligned. }
	\end{minipage}\hspace{3pc}%
	\begin{minipage}{16pc}
		\includegraphics[width=17pc]{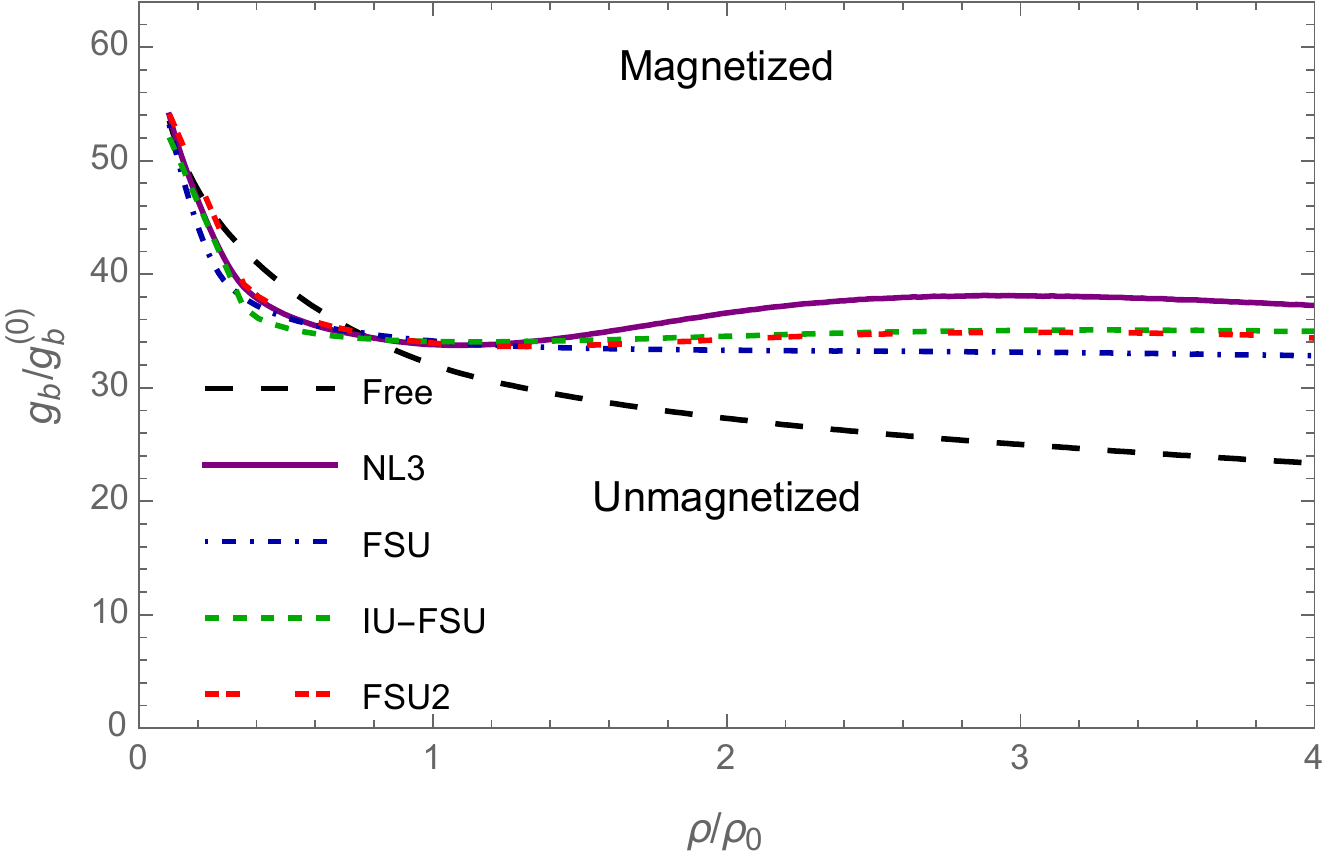}
		\caption{\label{fig:gbsym}Ferromagnetic phase boundary as multiples of the baryon dipole moment $g_b$ for free and interacting beta-equilibrated charge-neutral neutron-rich matter (including protons, neutrons and leptons) for nuclear densities as multiples of the nuclear saturation density $\rho_0$.}
	\end{minipage} 
\end{figure}

 In the case of $H=0$ then $P_{\perp}-P_{\parallel}  = 0$ and the anisotropy in the pressure disappears. While $H=0$ minimizes the Gibbs Free Energy of the system, it also implies that $B=M$, which by itself does not mean the system is magnetized.  However, if $B=M\neq0$, it does imply that the system is in a ferromagnetic state, with energy lower than the $B=M=0$ state \cite{DienerSpin}.\\
 \\
 One possible mechanism for achieving a magnetized $H=0$ state is if the density-dependence of the fermion dipole moment is such that it increases substantially with density.  However, such a density-dependence is only expected from the baryon dipole moment, due to these particles' underlying quark substructure \cite{DienerSpin}.  By increasing the baryon dipole moments, ferromagnetic phase boundary is reach at multiples of $\sim35$ times that of the normal dipole moments ($g_b^{(0)}$), as shown in figure \ref{fig:gbsym} where the ferromagnetic phase boundary is shown for an free beta-equilibrated charge-neutral neutron-rich gas as well as interacting nuclear matter systems under the same equilibrium conditions (details of the calculation are given in Ref. \cite{DienerSpin}). Note the consistency in the phase boundary for the FSU, IU-FSU, and FSU parameter sets, all of which were calibrated for nuclear studies related to neutron star properties \cite{IUFSU,FSU2}.
 %

 \section{Discussion}\label{sec:dis}
Based on the high multiples of the baryon dipole needed to cross the phase boundary, the ferromagnetic phase would be a high density effect, if achieved at all in dense matter systems.  However, when the phase boundary is crossed, $B$ is of the order of $10^{17}$ G \cite{DienerSpin}, corresponding to magnetic energies per particle of X- to $\gamma$-rays ($\sim$ keV and above). This is not unexpected, the magnetic field is of nuclear origin and thus the corresponding nuclear energy scale and since magnetars are observed as X- and $\gamma$-ray sources \cite{magobs, magnetars}. \\
\\
Nuclear ferromagnetism has been proposed and investigated in various studies (see \cite{DienerPhD} for a summary as well as the more recent \cite{Koji}), but is not widely considered in the literature to be the origin, nor major contributing factor to the superstrong magnetar magnetic field \cite{magobs, magnetars}. While the ferromagnetic field would lower the energy density of the system and thus soften the EoS of the neutron star, there is no drastic (observable) reduction in the maximum mass of the star is expected \cite{DienerPhD, Chat2015}.\\
\\
However, if the idea of a ferromagnetic phase transition can be entertained, there are some interesting \emph{possible} implications if such a transition where to take place in the star. Firstly, a transition to a ferromagnetic star would be accompanied by massive release of energy, in the range from soft X-rays to to $\gamma$-rays, due to the ferromagnetic state have lower energy from the normal or even externally magnetized state \cite{DienerPhD}. Given the high density as well as the volume of the star, the observed massive energy release/luminosities (energy release of $\sim 10^{35}$erg for one FRB \cite{FRBmag}) can also be accounted for\footnote{A quick estimate shows that an energy difference of the order of $1$ MeV/fm$^3$ corresponds to luminosities of the order of more than $10^{50}$ ergs/m$^3$. Of course such an energy is not realistic in terms of actually being observed/emitting by the star, but rather indicates that observed energy release is not out of range.}.\\
\\
 Secondly, since the star will be transitioning from a $H\neq 0$ state to a $H=0$ state, the anisotropy in the pressure will disappear. Also all high mass moments (arising from the star's deformation) will disappear as well, since the star will become spherical symmetric.  Thus the implication will be that if the magnetar's gravitational wave emission can be compared before and after the phase transition, the gravitational wave emission from the star will have ceased or be dramatically influenced post the phase transition.\\
 \\
  Thirdly, the change in the mass moment of the star, could also result in a sudden change in angular moment, changing the rotational period of the star.  Depending on how rapidly an associated reconfiguration of the system and distribution of the angular momentum takes place, this mechanism could possibly account for or contribute to glitching in magnetars. Correspondingly a change in the magentar's magnetic field during a glitch could also be indicative of a ferromagnetic phase contributing to glitching in magnetars Unfortunately the magnetar magnetic field is mostly inferred from its spin period and its derivative \cite{magnetars}, so there seems to currently be no independent/direct measure of the magnetar magnetic field. \\
  \\
  Fourth, due to the degeneracy of charged particle Landau-levels, charged particle fractions are higher when compared to unmagnetized systems \cite{DienerPhD}.  While the $H\neq0$ state is obviously magnetized, when transitioning to $H=0$ state a significant number of baryons could undergo inverse beta-decay, resulting in the release of many anti-neutrinos. \\
\\
It must be pointed out that the time scale and macroscopic mechanisms of the transition of the $H\neq 0$ state to a $H=0$ state is unknown and the subject of ongoing research. However, based on the above anecdotal arguments it is worth investigating the thermodynamic  behaviour of dense nuclear matter systems, as these investigations could prove fundamental in describing the observed properties of strongly magnetized systems like magnetars or even FRBs. 
%
%
%
\section{Conclusion}
While there is still much to be learned about magnetized dense nuclear matter and its behaviour, indications are that multi-wavelength observations of astrophysical objects could provide much insight into these systems. In this regard development of both terrestrial nuclear experiments and the new generation of telescope like the Square Kilometre Array (SKA) could reveal much about the very big as well as the very small. 
 \ack
 This research is supported by the BIUST Initiation Grant No.\ R00047.


\section*{References}
 \begin{thebibliography}{9}
 	\bibitem{annMR} \"{O}zel F and Freire P 2016 \textit{Annu. Rev. Astron. Astrophys.} {\bf 54}  401
 	\bibitem{annnuc} Hebeler K, Holt J D, Menendez J and Schwenk A 2015 \textit{Annu. Rev. Nucl. Part. Sci.} {\bf 65}  457
 	\bibitem{nsobs} Enoto T , Kisaka S and Shibata S 2019 \textit{Rep. Prog. Phys.} {\bf 82}  106901
 	\bibitem{magobs} Turolla R, Zane S and Watts A L  2015 \textit{Rep. Prog. Phys.} {\bf 78}  116901
 	\bibitem{magnetars} Kaspi V M and Beloborodov A M 2017 \textit{Annu. Rev. Astron. Astrophys.} {\bf 55}  261
 	\bibitem{FRBmag} Bochenek C D, Ravi V, Belov K V, Hallinan G, Kocz J, Kulkarni S R and McKenna D L 2020 \textit{Nature} \textbf{587} 59
 	\bibitem{DienerSpin}  Diener J P W and Scholtz F G 2020 \textit{Phys. Rev. C} \textbf{102}  055805
 	\bibitem{NL3} Lalazissis G A,  K\"{o}nig J, and  Ring P  1997  \textit{Phys. Rev. C} \textbf{55}   540
 	\bibitem{FSU} Todd-Rutel B G and Piekarewicz J 2005  \textit{Phys. Rev. Lett} \textbf{ 95}  122501
 	\bibitem{IUFSU}   Fattoyev F J, Horowitz  C J,   Piekarewicz J and Shen G  2010  \textit{Phys. Rev. C} \textbf{82}   055803
 	\bibitem{FSU2}  Chen W-C and Piekarewicz J 2014  \textit{Phys. Rev. C} \textbf{90}   044305
 	\bibitem{DienerPhD}  Diener J P W 2012  Ph.D. thesis Stellenbosch University (\textit{Preprint} nucl-th/1305.7346)
 	\bibitem{Brod00} Broderick A, Prakash M and Lattimer J M 2000 \textit{Astrophys. J.} \textbf{537} 351.
 	\bibitem{DienerVac}  Diener J P W and Scholtz F G 2020 \textit{Phys. Rev. C} \textbf{101}  035808
 	\bibitem{Bland82} Blandford R D and Hernquist L 1985\textit{ J. Phys. C} \textbf{15} 6233 
 	\bibitem{Ferrer2010} Ferrer E J, de la Incera V, Keith J P, Portillo I and  Springsteen P L 2010 \textit{Phys. Rev. C} \textbf{82} 065802
 	\bibitem{Strick} Strickland M, Dexheimer V and Menezes D P  2012 \textit{Phys. Rev. D} \textbf{86} 125032
 	\bibitem{Koji} Hashimoto K  2015 \textit{Phys. Rev. D} \textbf{19} 085013
 	\bibitem{Chat2015} Chatterjee D, Elghozi T, Novak J and Oertel M 2015 \textit{Mon. Not. R. Astron. Soc. }\textbf{447} 3785
 	
 	
 	
 	
 	
%
%
%
%
 \end{thebibliography}
\end{document}